\PassOptionsToPackage{table}{xcolor}
\documentclass[sigconf]{acmart}
\usepackage{multirow}
\usepackage{makecell}
\usepackage{subcaption}
\AtBeginDocument{%
  }

\makeatletter
\newcommand{\correspondingauthornote}[1]{%
  \if@ACM@anonymous\else
    \g@addto@macro\@authornotes{%
      \begingroup
        \def\@thefnmark{\ding{41}}%
        \@footnotetext{#1}%
      \endgroup}%
  \fi}
\makeatother

\copyrightyear{2026}
\acmYear{2026}
\setcopyright{cc}
\setcctype{by}
\acmDOI{10.1145/3767308.3836357}
\acmConference[MM '26]{Proceedings of the 34th ACM International Conference on Multimedia}{November 10--14, 2026}{Rio de Janeiro, Brazil}
\acmBooktitle{Proceedings of the 34th ACM International Conference on Multimedia (MM '26), November 10--14, 2026, Rio de Janeiro, Brazil}
\acmISBN{979-8-4007-2213-4/2026/11}
\begin{document}

\title{GVC-RT: Towards Real-Time Generative Video Compression at Ultra-Low Bitrates}

\author{Tianjian Dang}
\authornote{Tianjian Dang and Sixian Wang contributed equally to this work and share first authorship.}
\orcid{0009-0004-5794-566X}
\email{dtj@bupt.edu.cn}
\affiliation{%
  \institution{Beijing University of Posts and Telecommunications}
  \city{Beijing}
  \country{China}}

\author{Sixian Wang}
\authornotemark[1]
\orcid{0000-0002-0621-1285}
\email{sxwang@sjtu.edu.cn}
\affiliation{%
  \institution{Shanghai Jiao Tong University}
  \city{Shanghai}
  \country{China}}

\author{Lei Luo}
\orcid{0000-0002-7008-4276}
\email{luolei@cqupt.edu.cn}
\affiliation{%
  \institution{Chongqing University of Posts and Telecommunications}
  \city{Chongqing}
  \country{China}}

\author{Guo Lu}
\orcid{0000-0001-6951-0090}
\email{luguo2014@sjtu.edu.cn}
\affiliation{%
  \institution{Shanghai Jiao Tong University}
  \city{Shanghai}
  \country{China}}

\author{Jincheng Dai}
\correspondingauthor
\correspondingauthornote{Jincheng Dai is the corresponding author.}
\orcid{0000-0002-0310-568X}
\email{daijincheng@bupt.edu.cn}
\affiliation{%
  \institution{Beijing University of Posts and Telecommunications}
  \city{Beijing}
  \country{China}}

\renewcommand{\shortauthors}{Tianjian Dang, Sixian Wang, Lei Luo, Guo Lu, \& Jincheng Dai}

\begin{abstract}
Recent generative video codecs (GVCs) have achieved impressive reconstruction fidelity at ultra-low bitrates (< 0.02 bits per pixel) by compressing the tokens from generative tokenizers. However, existing GVCs generally require considerable computation time and model complexity, which hinder their deployment on compute-limited devices and in real-time applications. To bridge this gap, we systematically identify the computational bottlenecks and propose GVC-RT, which redesigns the generative latent coding framework to realize real-time video coding without sacrificing compression performance. Specifically, built on a pretrained lookup-free quantization (LFQ) tokenizer, GVC-RT adopts an asymmetric architecture that directly learns to match the LFQ latent distribution, while generative-space alignment is enforced via a regularization loss term only during training. In this manner, we bypass heavy tokenization and entirely remove the complex feature-alignment process at inference time. Moreover, we further introduce a lightweight de-tokenizer architecture to resolve the final latency bottleneck during decoding. Experimental results demonstrate that GVC-RT outperforms the previous SOTA model, GLC-Video, with average BD-rate savings of 12.4\% and 48.8\% in terms of DISTS and LPIPS, while achieving encoding/decoding speeds of 123.1/55.1 fps for 1080p video. The code is at \url{https://github.com/semcomm/GVC-RT}.
\end{abstract}



\maketitle

\section{Introduction}

\begin{figure}[t]
  \centering
    \includegraphics[width=\linewidth]{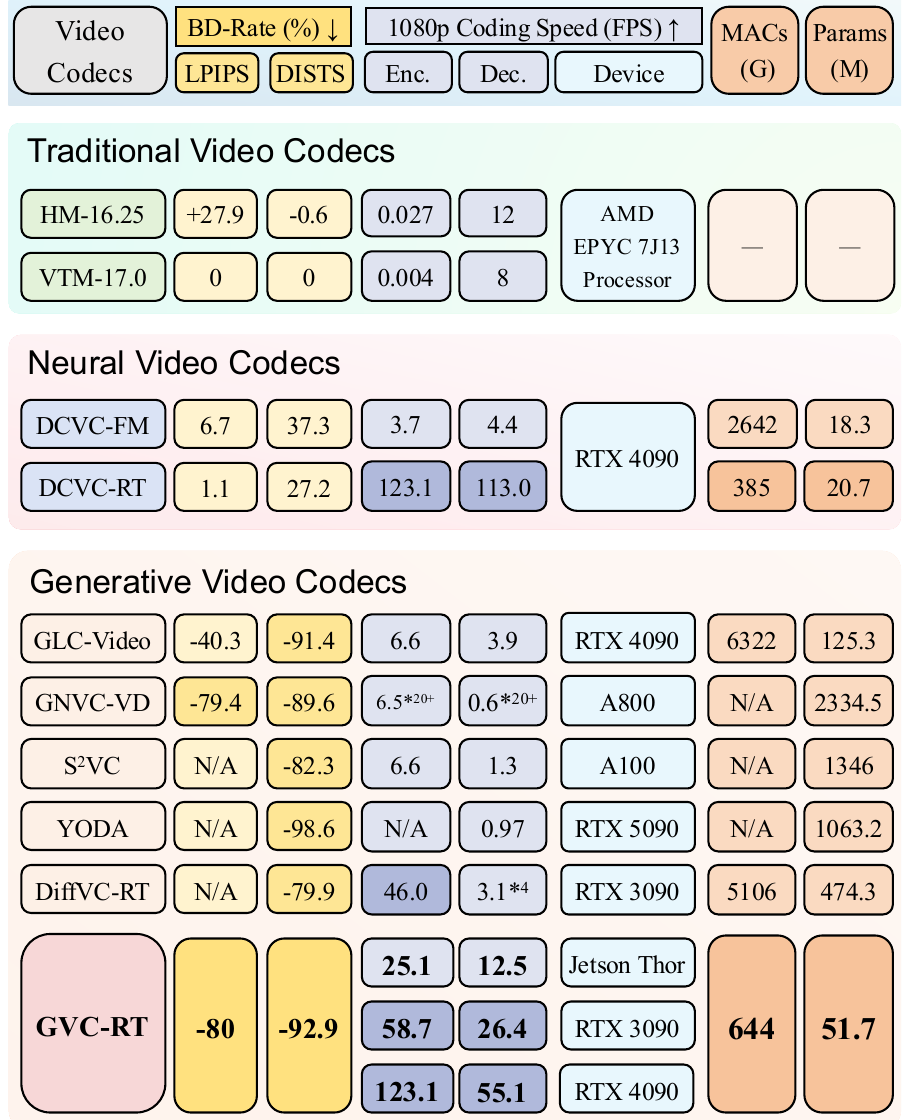}
  \caption{Coding efficiency and runtime comparison. BD-Rate is evaluated on the 1080p UVG dataset using VTM-17.0 as the anchor. Here, ``*$^n$'' indicates that the reported encoding/decoding speed is amortized over a temporal window of $n$ frames; ``--'' indicates that the metric is not applicable to this type of codec; and ``N/A'' means that the code and data are not provided in the original publication.}
  \Description{A comparison of traditional, neural, and generative video codecs by LPIPS and DISTS BD-Rate, 1080p encoding and decoding speed, computation, and model size. GVC-RT combines strongly negative perceptual BD-Rates with real-time speed on RTX 3090 and RTX 4090 GPUs.}
  \label{fig:real_time_compare}
\end{figure}

The exponential growth of visual data poses persistent challenges for bandwidth and storage, driving an ever-increasing demand for highly efficient video compression. 
As the performance gains of traditional handcrafted codecs gradually saturate, neural video compression (NVC) has emerged as a promising alternative. 
By leveraging data-driven spatial-temporal modeling, recent NVC methods~\cite{Lu_2019_CVPR,li2021dcvc,DCVCDC,fvc,dcvcfm,dhvc,DCVCRT} have achieved continuous advances in rate-distortion (RD) performance. 
Notably, practical neural codecs like DCVC-RT~\cite{DCVCRT} outperform the latest traditional standard H.266/VVC~\cite{vvc} while maintaining real-time processing speeds.

However, superior RD performance does not inherently guarantee high visual quality~\cite{GLCvideo}. 
The widely used distortion objectives (e.g., MSE and MS-SSIM~\cite{MS-SSIM}) correlate weakly with human perception. 
This discrepancy becomes particularly pronounced at ultra-low bitrates (typically < 0.02 bits per pixel), where RD-optimized codecs suffer from severe perceptual degradations such as blocking artifacts and oversmoothed textures, leading to a sharp decline in quality of experience (QoE).

To preserve perceptual quality under extreme compression, recent research has turned to generative video compression (GVC). 
By incorporating powerful generative priors and optimizing for perceptual objectives, GVC methods~\cite{GLCvideo,yoda,gnvc,s2vc,DiffVC,DiffVC_RT} have demonstrated the capability to synthesize realistic textures and preserve high-fidelity details at ultra-low bitrates. 
However, their success is largely attributed to the adoption of complex pretrained encoders and powerful diffusion or transformer decoders~\cite{GLCvideo,yoda,gnvc,s2vc,DiffVC}, which incur massive computational overhead and severe latency, hindering deployment on compute-limited devices and in real-time applications. 
Specifically, existing methods typically follow a generative latent coding (GLC) paradigm~\cite{GLCvideo}, relying on heavy generative tokenizers to transform the coding space into a generative latent space aligned with human perception, so that token compression distortions can be effectively compensated at the receiver with the help of pretrained or finetuned generative priors (e.g., diffusion models). 
Despite achieving superior rate-distortion-perception (RDP) performance, this paradigm necessitates computationally expensive tokenization and complex feature alignment during inference~\cite{yoda,gnvc,s2vc,DiffVC}. 
In contrast to MSE-optimized codecs such as DCVC-RT~\cite{DCVCRT} that readily support real-time processing, current GVCs remain far from practical deployment. 

To bridge this gap, we propose a novel real-time generative compression framework named GVC-RT, which systematically redesigns the GLC pipeline to realize real-time video coding without sacrificing perceptual performance. 
We first analyze the module-wise latency breakdown of the standard GLC paradigm and identify three primary inference bottlenecks: heavy tokenization, complex feature alignment, and costly de-tokenization. 
To eliminate these bottlenecks, we streamline the pipeline through a threefold strategy. 
First, inspired by~\cite{zhang2025ultra}, we adopt an asymmetric architecture that bypasses explicit tokenization, instead employing the lightweight DCVC-RT encoder~\cite{DCVCRT} to directly learn to extract the information essential for downstream generation. 
Second, to resolve the feature alignment bottleneck, we train the compression network to reconstruct discrete tokens by minimizing the cosine distance to the codewords of the pretrained lookup-free quantization (LFQ) tokenizer~\cite{FSQ,LMBeatsDiff,luo2024open}. This generative space alignment is applied only during training, which bypasses complex alignment modules at inference time. 
Third, we distill the original LFQ de-tokenizer into a lightweight variant, which resolves the final latency bottleneck for real-time decoding.

Benefiting from these structural innovations, GVC-RT achieves a favorable balance among coding speed, compression efficiency, and reconstruction fidelity.
As shown in Fig.~\ref{fig:real_time_compare}, it supports real-time 1080p compression on a single consumer-grade NVIDIA RTX 4090 GPU, achieving 123.1 fps for encoding and 55.1 fps for decoding.
Meanwhile, compared with concurrent works~\cite{yoda,gnvc,s2vc,DiffVC_RT}, GVC-RT maintains highly competitive perceptual quality.
Notably, relative to DiffVC-RT~\cite{DiffVC_RT}, which is also designed for real-time deployment, GVC-RT runs substantially faster even without using multi-frame parallel decoding strategy~\cite{DiffVC_RT}, while delivering better rate-distortion-perception performance.
To the best of our knowledge, GVC-RT represents an early step toward practical real-time deployment of generative codecs on consumer-grade hardware.

In summary, the main contributions of this paper are as follows:

\begin{itemize}
\item We propose a novel generative video compression framework, termed GVC-RT, for practical real-time ultra-low bitrate coding. Motivated by the inefficiency of existing generative latent coding pipelines, the proposed framework eliminates the heavy inference-time tokenization and complex generative alignment modules at the encoder side by directly learning generative-aware compression with a lightweight DCVC-RT-based encoder.

\item We further redesign the decoding pipeline for efficient real-time generation. Specifically, GVC-RT shifts the generative-space alignment entirely to training through LFQ codeword supervision, and replaces the original heavy de-tokenizer with a distilled lightweight variant, thereby substantially reducing decoding latency while preserving the perceptual benefits of generative priors.

\item Extensive experiments show that GVC-RT achieves a strong balance among perceptual quality, reconstruction fidelity, and coding efficiency at ultra-low bitrates. In particular, it enables real-time 1080p compression on a single consumer-grade RTX 4090 GPU at 123.1 fps encoding and 55.1 fps decoding, while maintaining highly competitive rate-distortion-perception performance compared with existing generative codecs.
\end{itemize}

\section{Related Work and Background}

\begin{figure*}[t]
  \centering
    \includegraphics[width=\linewidth]{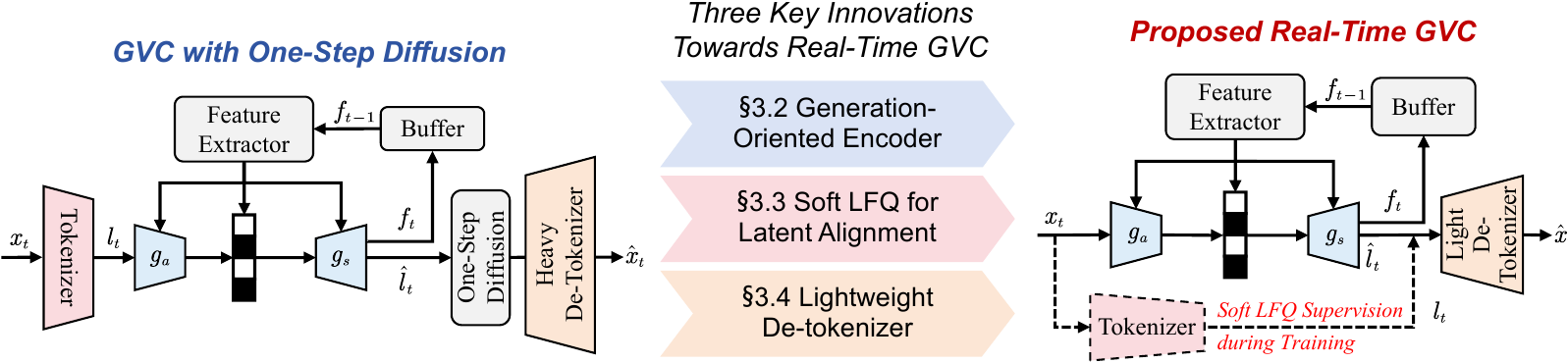}
    \caption{Framework redesign. Left: a GVC with one-step diffusion (GVC-OSD) built on the GLC paradigm. Right: our proposed GVC-RT framework. To enhance efficiency, we establish a generation-oriented encoder, utilize the soft LFQ for latent alignment, and distill an efficient de-tokenizer.
} 
  \Description{Side-by-side diagrams of the GVC-OSD and GVC-RT pipelines. The redesigned GVC-RT pipeline removes inference-time tokenization and complex alignment, trains a generation-oriented encoder with soft LFQ alignment, and replaces the original de-tokenizer with a distilled lightweight version.}
  \label{fig:paradigm}
\end{figure*}

\subsection{Generative Video Compression}

Neural video compression (NVC) achieves competitive RD performance against traditional standards by leveraging non-linear transforms and end-to-end RD optimization. Early explorations, pioneered by DVC~\cite{Lu_2019_CVPR}, primarily relied on optical flow estimation and residual coding architectures. 
Subsequent works~\cite{li2021dcvc, dhvc, dcvcfm} have explored conditional coding by mining diverse spatial-temporal contexts, further improving RD performance. 
While these MSE-optimized NVCs excel at regular bitrates, they tend to produce overly smoothed reconstructions at ultra-low bitrates, where the available bits are insufficient to maintain pixel-level fidelity.

To overcome this limitation, leveraging generative priors to synthesize realistic textures and mitigate artifacts becomes a natural strategy. Early efforts such as PLVC~\cite{PLVC} explored adversarial training for this purpose. More recently, the generative latent coding (GLC) paradigm~\cite{GLCvideo} has emerged as the dominant approach, achieving high-fidelity reconstruction by compressing discrete tokens extracted via pretrained generative tokenizers. To further push the perceptual upper bound, diffusion models~\cite{diffusion} have been incorporated into the compression pipeline as generative decoders.
However, while recent methods~\cite{yoda,s2vc,DiffVC_RT} adopt one-step generation to bypass the iterative denoising latency, the overall framework is still bottlenecked by heavy tokenizers and large decoders, preventing real-time processing.

\subsection{Real-Time Neural Video Coding}

As discussed, accelerating generative video compression remains highly challenging due to the substantial computational overhead of the generation modules. To address real-time constraints, valuable insights can be drawn from existing acceleration efforts in neural video codecs. Specifically, DCVC-RT~\cite{DCVCRT} establishes an efficient structural paradigm by operating exclusively on a single low-resolution latent representation alongside implicit temporal modeling. By mitigating operational complexity (e.g., frequent memory access), it achieves exceptional real-time efficiency. More directly related to our work, the concurrent DiffVC-RT~\cite{DiffVC_RT} explores real-time generative coding. However, its real-time capability relies heavily on high-end professional accelerators (e.g., NVIDIA H800 GPUs) to execute the computationally expensive denoising processes. Furthermore, its use of multi-frame parallel decoding inherently introduces a fixed structural delay of several frames, precluding its use in low-latency settings. 
As a result, achieving real-time generative coding under ultra-low bitrate conditions remains largely unexplored, mainly due to the inherent computational overhead of generative models.

\begin{figure}[t]
  \centering
    \includegraphics[width=0.92\linewidth]{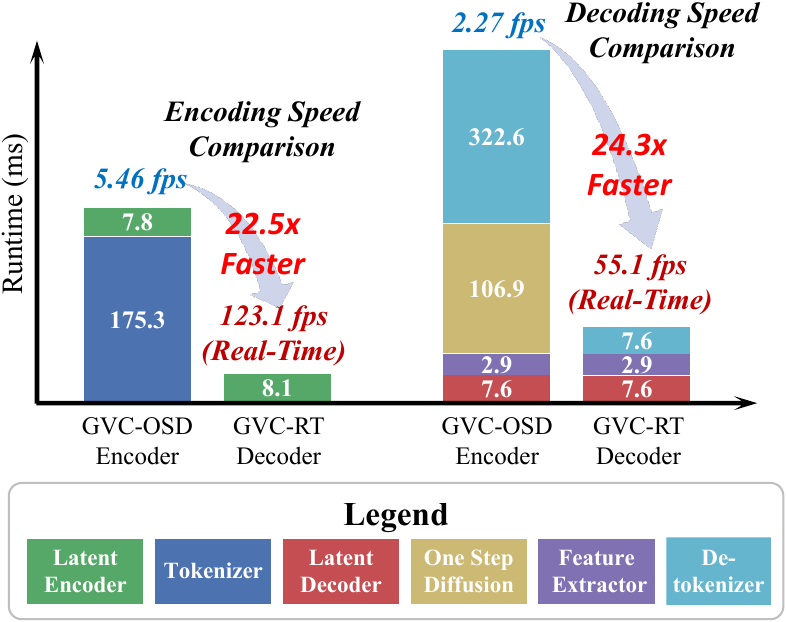}
    \caption{Inference latency of the individual modules in GVC-OSD and GVC-RT.} 
  \Description{A module-by-module latency comparison showing that GVC-RT reduces or removes the dominant tokenization, feature-alignment, and de-tokenization costs of GVC-OSD.}
  \label{fig:latency}
\end{figure}

\section{The Proposed GVC-RT}

\begin{figure*}[t]
  \centering
    \includegraphics[width=0.92\linewidth]{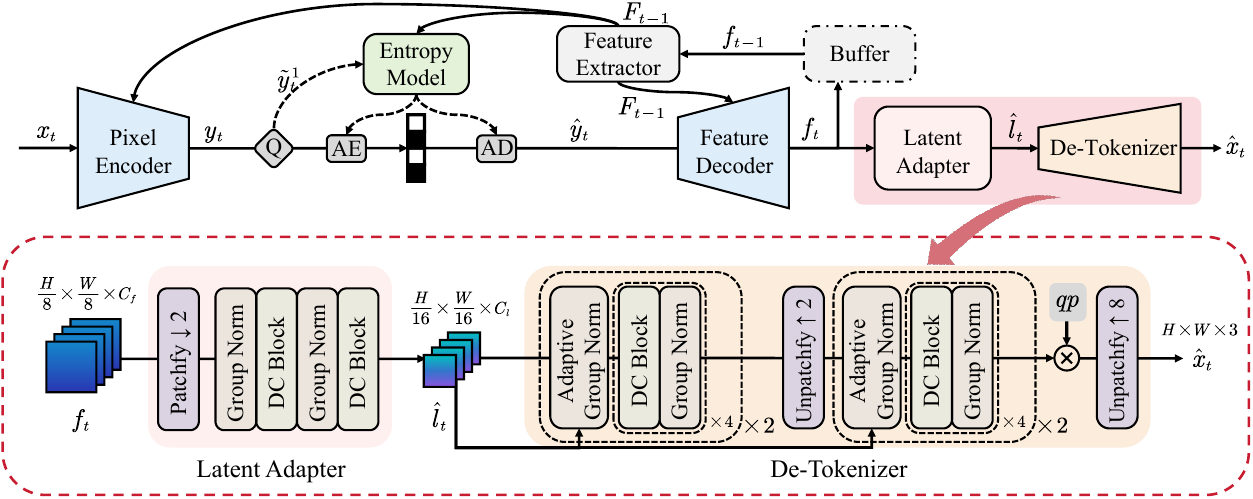}
    \caption{Framework overview. DC Block, Q, AE and AD represent depth-wise convolution block~\cite{DCB}, quantization, arithmetic encoder and decoder, respectively. The structures of the Latent Adapter and De-tokenizer are shown in detail.} 
  \Description{The GVC-RT encoder-decoder architecture and training connections, including depth-wise convolution blocks, quantization, arithmetic coding, temporal context, the latent adapter, and the lightweight de-tokenizer. Insets detail the latent-adapter and de-tokenizer layers.}
  \label{fig:framework}
\end{figure*}

\subsection{Motivation and Bottleneck Analysis}\label{sec:motivation}

While GVCs achieve superior perceptual quality at ultra-low bitrates, their computational overhead remains a critical barrier to real-time deployment. Existing SOTA GVCs, including accelerated one-step diffusion models~\cite{DiffVC_OSD,yoda,s2vc}, typically adapt to the GLC paradigm~\cite{GLCvideo}.

To quantitatively investigate real-time GVC deployment, we construct a representative baseline, GVC-OSD. For fair evaluation, we adopt the DCVC-RT latent codec~\cite{DCVCRT} for latent compression and standard tokenizer and de-tokenizer architectures, together with a latent-alignment diffusion model implemented using Diffusers~\cite{diffusers}. Module-wise latency profiling on an RTX 4090 GPU (Fig.~\ref{fig:latency}) reveals three critical bottlenecks:

\noindent\textbf{Heavy Tokenization:} Explicitly mapping input frames into a generative latent space consumes the majority of encoding time.

\noindent\textbf{Complex Generative Alignment:} Forcing compressed features to align with generative priors at the decoder side introduces substantial latency.

\noindent\textbf{Costly De-Tokenization:} The final spatial reconstruction via heavily parameterized decoders accounts for the most significant portion of the decoding latency.

To overcome these bottlenecks, GVC-RT systematically redesigns the encoding, alignment, and decoding stages to achieve real-time efficiency without compromising generative performance.

\subsection{Framework of GVC-RT}

As illustrated in Fig.~\ref{fig:framework}, given an input frame $x_t \in \mathbb{R}^{h \times w \times 3}$ at time step $t$ and the temporal context feature $F_{t-1}$ extracted from the last decoded frame, our asymmetric encoder $E(\cdot, \cdot)$, adapted from the DCVC-RT encoder~\cite{DCVCRT}, maps the current frame into a latent representation $y_t$ as formulated in Eq.~\eqref{eq:encoder_mapping}:
\begin{equation}
y_t = E(x_t, F_{t-1})
\label{eq:encoder_mapping}
\end{equation}
The latent $y_t$ is then quantized into $\hat{y}_t$ and entropy-coded with a learned probabilistic model into the bitstream. 

On the decoder side, the reconstructed latent $\hat{y}_t$ is first decoded into the frame feature $\hat{f}_t$ by the feature decoder $D_F(\cdot, \cdot)$ from the DCVC-RT decoder~\cite{DCVCRT}, using the temporal context feature $F_{t-1}$ as shown in Eq.~\eqref{eq:feature_decoder}:
\begin{equation}
\hat{f}_t = D_F(\hat{y}_t, F_{t-1})
\label{eq:feature_decoder}
\end{equation}
The feature extractor then extracts the temporal context feature $F_{t}$ for the next frame.

The latent adapter then transforms $\hat{f}_t$ into the aligned latent $\hat{l}_t$. The reconstructed frame $\hat{x}_t$ is generated by the de-tokenizer $D_T(\cdot)$ in Eq.~\eqref{eq:frame_reconstruction}:
\begin{equation}
\hat{x}_t = D_T(\hat{l}_t)
\label{eq:frame_reconstruction}
\end{equation}
This pipeline integrates a real-time NVC backbone~\cite{DCVCRT} with a lightweight de-tokenizer, achieving perceptually faithful and real-time inference under extreme compression.

\begin{figure}[t]
  \centering
    \includegraphics[width=1\linewidth]{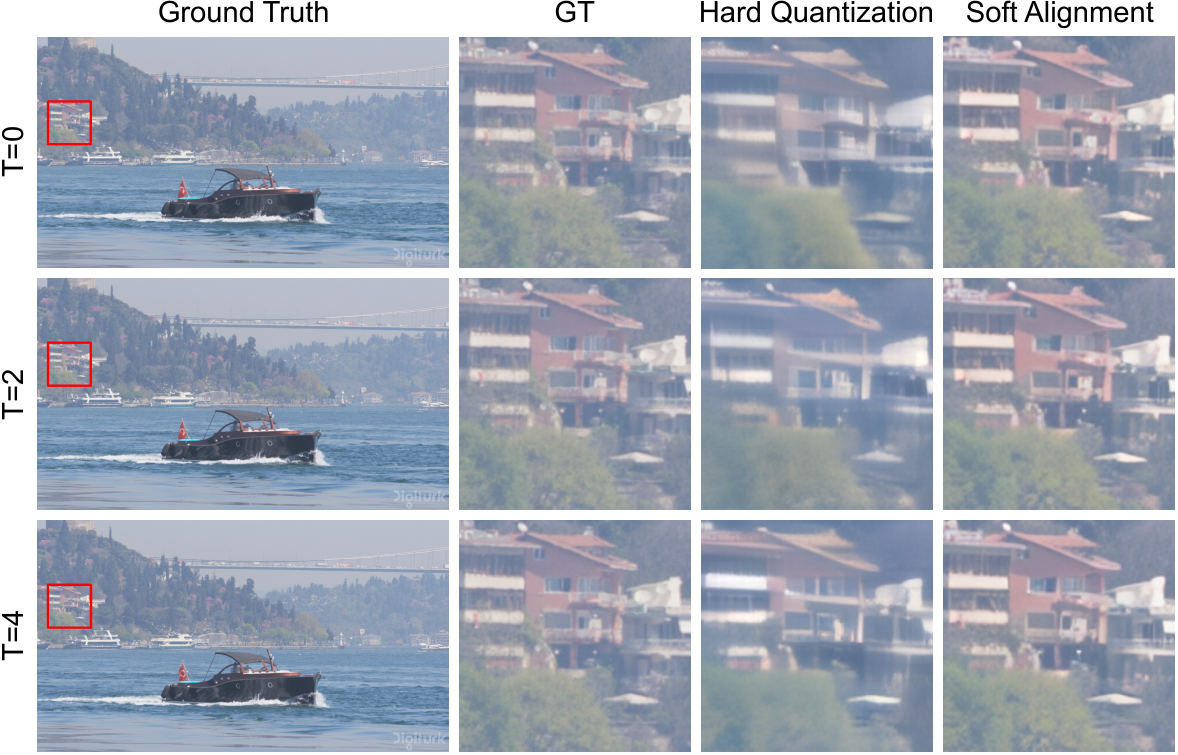}
    \caption{Comparison of hard quantization and soft alignment of LFQ. The hard quantization leads to severe temporal flickering artifacts due to sign flips, while the soft alignment mitigates it by learning the continuous latent.} 
  \Description{Video-frame crops and latent trajectories compare hard LFQ quantization with soft LFQ alignment. Hard sign changes produce visible temporal flicker, whereas continuous soft alignment yields stable reconstructed content.}
  \label{fig:hard_vs_soft}
\end{figure}

\subsection{Asymmetric Generation-Aware Encoder}

To eliminate the latency introduced by explicit tokenization at the encoder side, we propose an asymmetric, generation-aware encoding architecture. In GLC pipelines~\cite{GLCvideo,s2vc}, the input frame is typically processed by a heavy tokenizer, mapping the image into a high-dimensional generative latent space, followed by a compression network to reduce spatial redundancy. 

Inspired by recent findings in generative image coding~\cite{zhang2025ultra}, we utilize the lightweight encoder from DCVC-RT~\cite{DCVCRT} to implicitly learn to extract the information required for downstream generative reconstruction, bypassing the explicit tokenization step entirely. 

By offloading the burden of generative space projection entirely to the decoder (details in Sec.~\ref{sec:alignment}), this asymmetric design reduces the encoding complexity. Concurrently, driven by the end-to-end generative supervision, $E(\cdot, \cdot)$ learns to optimally preserve the semantic and textural priors necessary for the subsequent high-realism synthesis.

\subsection{Soft LFQ for Generative Space Alignment}\label{sec:alignment}

As discussed in Sec.~\ref{sec:motivation}, relying on explicit generative alignment diffusion at the decoder side introduces significant latency. Since our asymmetric encoder extracts features without explicit tokenization, the decompressed latents at the decoder side must be aligned with a generative space to ensure high-realism synthesis. To achieve real-time decoding, we propose shifting the burden of latent alignment entirely from the inference phase to the training phase via a soft regularization strategy.

However, as illustrated in Fig.~\ref{fig:hard_vs_soft}, the hard-quantized discrete tokens $l_{tq}$ introduce temporal inconsistencies in the video domain. Slight inter-frame variations cause lossy-compressed values in $l_t$ near the zero boundary to oscillate, leading to abrupt sign flips in $l_{tq}$ and severe temporal flickering artifacts.

To retain the capacity of LFQ while maintaining temporal stability, we propose a Soft Alignment mechanism. The decompressed latent $\hat{l}_t$ directly learns the distribution of $l_t$, rather than the binary tokens $l_{tq}$, through the loss in Eq.~\eqref{eq:soft_alignment}:
\begin{equation}
\mathcal{L}_{cos}(\hat{l}_t, l_t),
\label{eq:soft_alignment}
\end{equation}
where $\mathcal{L}_{cos}(\cdot, \cdot)$ measures the cosine similarity between the reconstructed latent $\hat{l}_t$ and the original latent $l_t$. 
Because the magnitude of $l_t$ inherently maintains structural confidence, this continuous regression provides smooth temporal transitions, completely avoiding the hard boundary. Crucially, by enforcing this soft alignment with the regularization loss $\mathcal{L}_{cos}$ during training, our network naturally projects the latent $\hat{l}_t$ into the LFQ generative manifold. This zero-cost inference-time alignment eliminates the need for computationally expensive alignment diffusion at the decoder side.

\begin{figure*}[t] 
  \centering
    \includegraphics[width=0.93\linewidth]{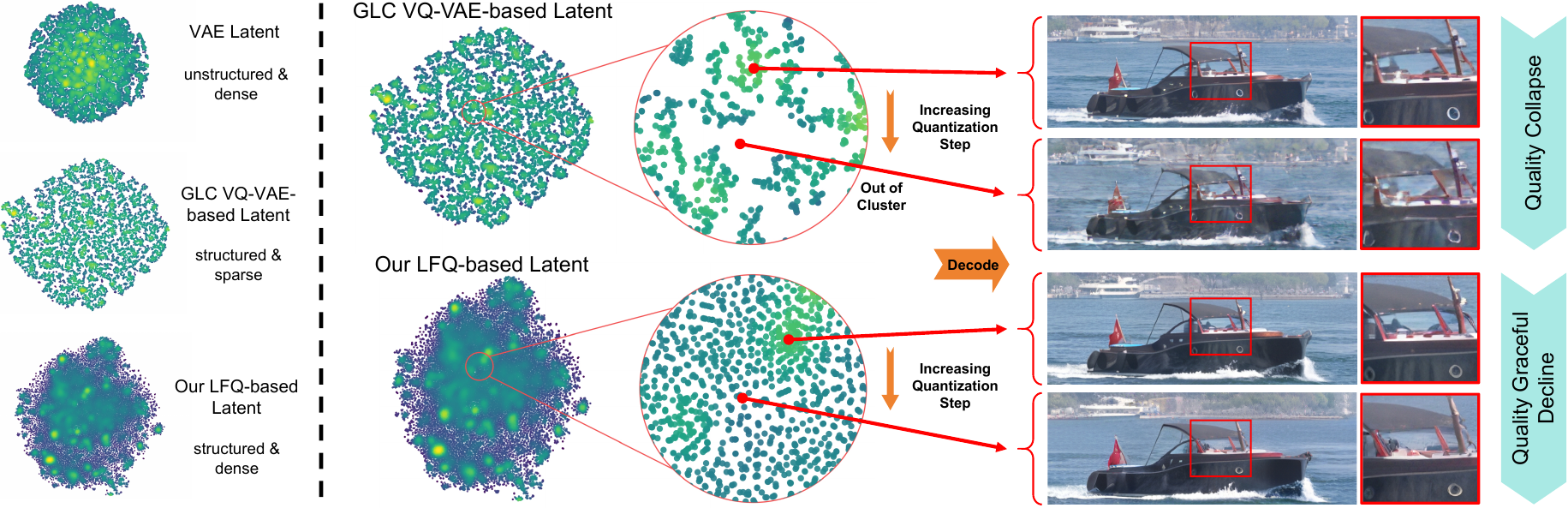}
    \caption{Left: visualization by t-SNE for comparing the latent distribution. Right: the impact of further compression of the latent space on frame generation.} 
  \Description{On the left, t-SNE plots compare the learned coding latents with the target LFQ latent distribution. On the right, reconstructed frame examples illustrate how further latent compression affects generated visual details.}
  \label{fig:stage2}
\end{figure*}

\subsection{Lightweight De-tokenizer with Latent-to-Image Distillation}\label{sec:detokenizer}

Even after eliminating explicit tokenization and inference-time alignment, the costly de-tokenizer remains the final bottleneck for real-time decoding. To resolve this, we redesign the spatial reconstruction stage by replacing the computationally expensive LFQ de-tokenizer with a highly efficient variant.

Architecturally, inspired by~\cite{DCVCRT}, we shift the computational burden to a $1/8$ downsampled latent space and replace standard ResBlocks with hardware-friendly depth-wise separable convolutions. 

However, significantly reducing network parameters inevitably compromises generative capacity. To maintain high-realism synthesis, we propose a latent-to-image distillation strategy. Instead of unstable end-to-end training from scratch, we freeze the pre-trained heavy LFQ tokenizer as the teacher. Given the LFQ codeword, our lightweight student de-tokenizer learns to reconstruct the image from the teacher-provided latent representation. This latent-to-image mapping effectively transfers the rich generative capabilities to our real-time decoder, ultimately overcoming the latency bottlenecks shown in Fig.~\ref{fig:latency}.

\section{Progressive Training of GVC-RT}

To ensure training stability and performance, we introduce a progressive three-stage training strategy.

\subsection{Stage I: Asymmetric LFQ-AE Adaptation}\label{sec:stage1}

In the initial stage, we aim to train the lightweight de-tokenizer proposed in Sec.~\ref{sec:detokenizer}. Direct end-to-end training of a lightweight generator from scratch is notoriously unstable. Therefore, we build upon the pre-trained LFQ-AE~\cite{luo2024open} to perform latent-to-image distillation.

Specifically, we freeze the heavy pre-trained LFQ encoder to provide a stable, highly structured continuous latent space $l_t$. To maintain feature consistency, our training objective $\mathcal{L}_{\text{Stage I}}$ follows the original LFQ's pre-training scheme and combines a reconstruction-perceptual loss with an adversarial loss, as defined in Eq.~\eqref{eq:stage1_loss}:
\begin{equation}
\mathcal{L}_{\text {Stage I}} = \mathcal{L}_{rec} + w_{adv} \cdot \mathcal{L}_{adv}
\label{eq:stage1_loss}
\end{equation}
where $w_{adv}$ is the weight assigned to the Patch-GAN adversarial loss~\cite{patchgan} $\mathcal{L}_{adv}$. The reconstruction-perceptual loss $\mathcal{L}_{rec}$ combines pixel-level and perceptual constraints, as shown in Eq.~\eqref{eq:reconstruction_loss}:
\begin{equation}
\mathcal{L}_{rec} = \|x - \hat{x}\|_1 + \mathcal{L}_{LPIPS}(x, \hat{x})
\label{eq:reconstruction_loss}
\end{equation}
Here, $\mathcal{L}_{LPIPS}$ represents the perceptual loss computed using LPIPS~\cite{lpips}, which leverages VGG features~\cite{vgg} to measure perceptual similarity. Additionally, the latent space provided by the frozen pre-trained encoder significantly reduces training costs while ensuring optimization stability.

\subsection{Stage II: Direct LFQ Latent Distribution Learning}

To directly learn the latent distribution, we decouple the asymmetric LFQ-AE optimized in Stage I into two components: a tokenizer and a de-tokenizer. The tokenizer acts as a latent extractor, mapping input frames into target LFQ representations to supervise the NVC compression pipeline (from the latent encoder to the latent decoder). Meanwhile, the de-tokenizer is loaded and frozen in the decoder. Crucially, to mitigate temporal flickering artifacts caused by sign flips, we utilize the continuous, pre-quantization LFQ latent as our supervision targets. Consequently, the optimization objective for Stage II is formulated in Eq.~\eqref{eq:stage2_loss}:
\begin{equation}
\mathcal{L}_{\text {Stage II}} = R + \lambda \cdot (\mathcal{L}_{rec} + \mathcal{L}_{distribution})
\label{eq:stage2_loss}
\end{equation}
The distribution alignment loss $\mathcal{L}_{distribution}$ ensures the reconstructed latent matches the original distribution, as defined in Eq.~\eqref{eq:distribution_loss}:
\begin{equation}
\mathcal{L}_{distribution} = w_{cos} \cdot \mathcal{L}_{cos}(l_t,\hat{l}_t) + w_{margin} \cdot \mathcal{L}_{margin}
\label{eq:distribution_loss}
\end{equation}
where $\mathcal{L}_{margin}$ enforces a structural constraint to push each dimension of $\hat{l}$ away from the semantically ambiguous intermediate regions. $w_{cos}$ and $w_{margin}$ are the weights assigned to the cosine similarity and marginal loss, respectively.

\subsection{Stage III: Joint Fine-tuning}

In the final stage, the entire network is jointly fine-tuned end to end to maximize the overall compression performance. Specifically, we unfreeze the parameters of the de-tokenizer, fully integrating it into the optimization pipeline. Furthermore, we incorporate adversarial loss to further improve the perceptual realism and high-frequency details of the reconstructed frames. Consequently, the overall optimization objective for Stage III is formulated in Eq.~\eqref{eq:stage3_loss}:
\begin{equation}
\mathcal{L}_{\text {Stage III}} = R + \lambda \cdot (\mathcal{L}_{rec} + w_{adv} \cdot \mathcal{L}_{adv} + \mathcal{L}_{distribution})
\label{eq:stage3_loss}
\end{equation}
\section{Perception-Driven Asymmetric Latent Supervision}\label{sec:discussion}

\begin{figure}[t]
  \centering
  \includegraphics[width=\linewidth]{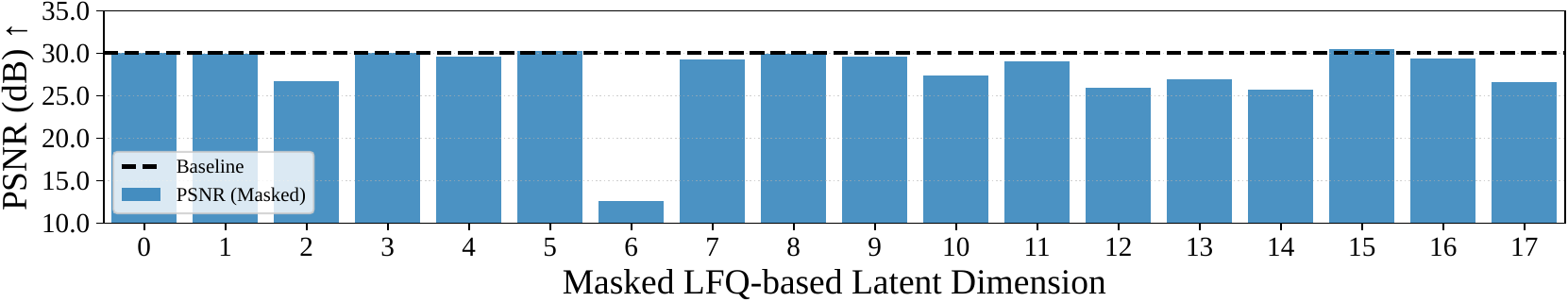}
  \includegraphics[width=\linewidth]{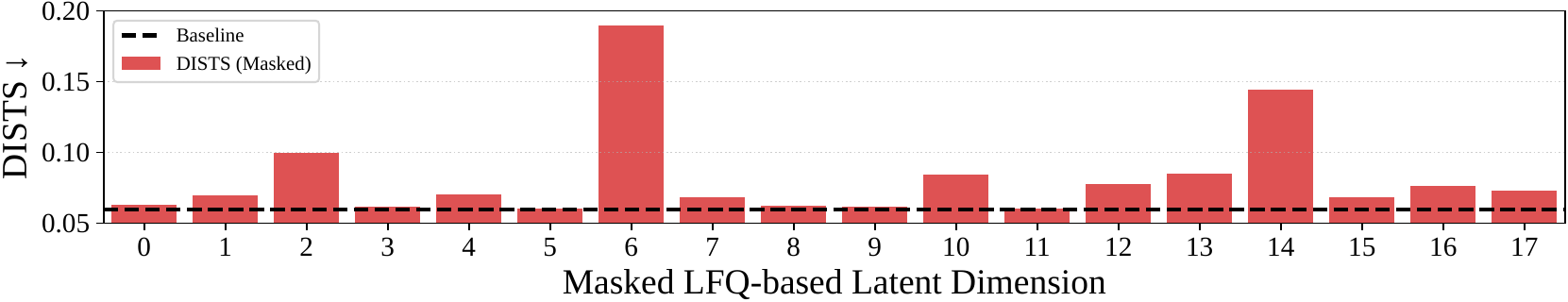}
  \caption{Impact of independently masking each LFQ dimension on reconstruction quality. The dashed lines represent the baseline performance without any masking.} 
  \Description{Two dimension-wise plots report the PSNR and DISTS changes caused by masking each LFQ latent dimension independently. Dashed horizontal lines mark the unmasked baseline, and the effects vary substantially across dimensions.}
  \label{fig:ablation_dim}
\end{figure}

To prevent generative reconstruction quality collapse at extremely low bitrates ($\text{bpp} < 0.005$) as shown in Fig.~\ref{fig:stage2}, we propose a perception-driven training strategy. Prior methods like GLC-Video~\cite{GLCvideo} apply homogeneous losses (e.g., MSE) uniformly across all latent channels. However, adapting this to our LFQ latents is suboptimal and leads to severe visual degradation (Sec.~\ref{sec:results} Ablation). As shown in Fig.~\ref{fig:ablation_dim}, the decoupled channels of LFQ contribute asymmetrically to generative quality: certain channels dictate critical semantics, while others control fine textures.

Driven by this insight, we abandon uniform supervision and introduce a dual-level paradigm:

\begin{itemize}
\item \textbf{Directional Latent Alignment:} We utilize Cosine Similarity instead of MSE, prioritizing the overall directional alignment of the high-dimensional latent space over rigid magnitude matching.
\item \textbf{Implicit Dimension Weighting:} By freezing the de-tokenizer during this stage, we introduce an image-domain perceptual loss. This forces the model to implicitly learn the varying importance of different latent dimensions directly from the reconstructed frame.
\end{itemize}

\noindent\textbf{Error-Attenuation Mechanism:} As shown on the right of Fig.~\ref{fig:stage2}, combined with our pre-quantization latent space, this strategy naturally attenuates errors. In ultra-low bitrate scenarios, less critical dimensions with high uncertainty are pushed to low numerical magnitudes (near zero). Consequently, even when inevitable sign flips occur, their impact on the continuous de-tokenizer is strictly suppressed, ensuring robust perceptual reconstruction under extreme bandwidth constraints.

\begin{figure*}[t]
  \centering
  \begin{subfigure}{\textwidth}
    \centering
    \includegraphics[width=0.75\textwidth]{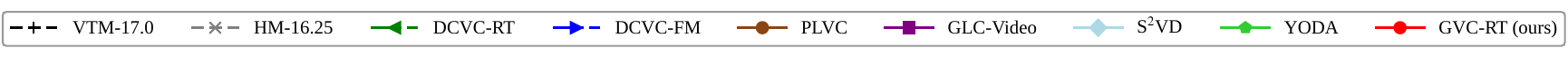}
  \end{subfigure}

  \begin{subfigure}{0.24\textwidth}
    \centering
    \includegraphics[width=\textwidth]{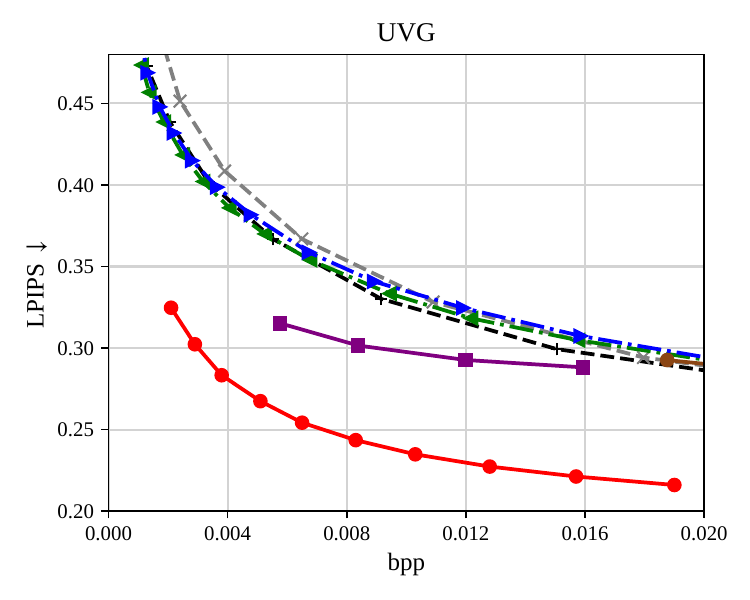}
  \end{subfigure}
  \begin{subfigure}{0.24\textwidth}
    \centering
    \includegraphics[width=\textwidth]{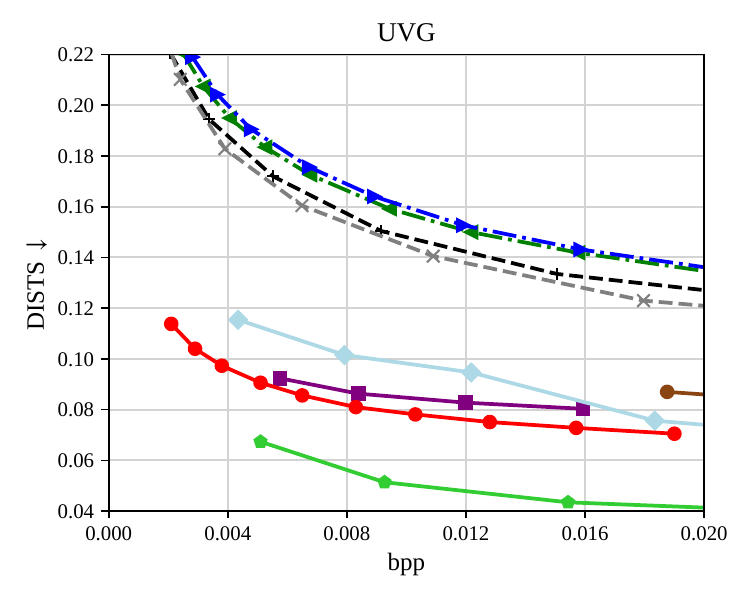}
  \end{subfigure}
  \begin{subfigure}{0.24\textwidth}
    \centering
    \includegraphics[width=\textwidth]{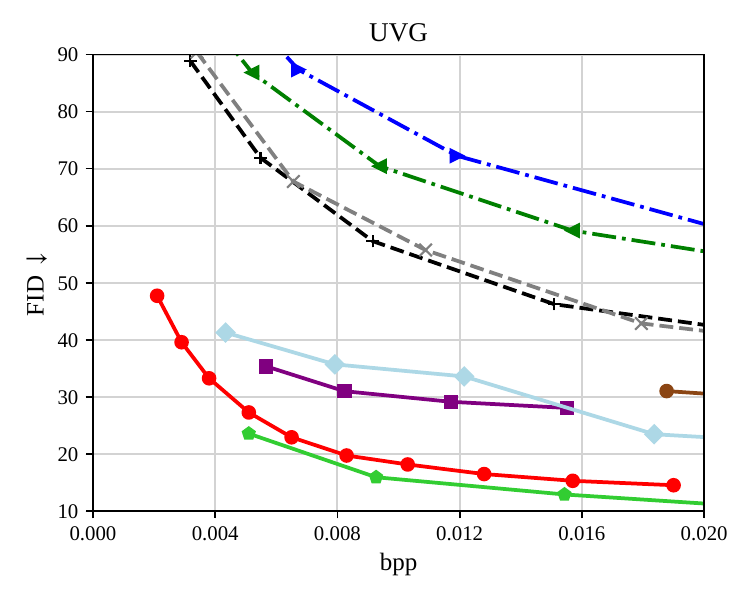}
  \end{subfigure}
  \begin{subfigure}{0.24\textwidth}
    \centering
    \includegraphics[width=\textwidth]{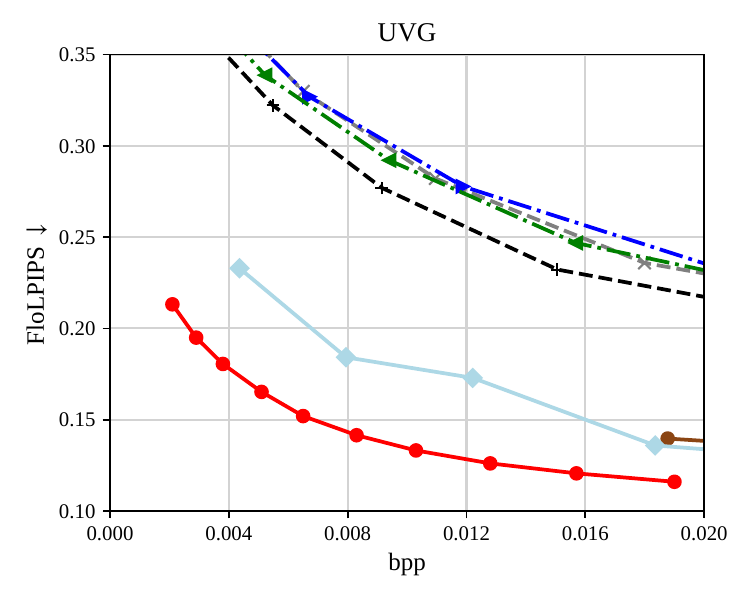}
  \end{subfigure}

  \begin{subfigure}{0.24\textwidth}
    \centering
    \includegraphics[width=\textwidth]{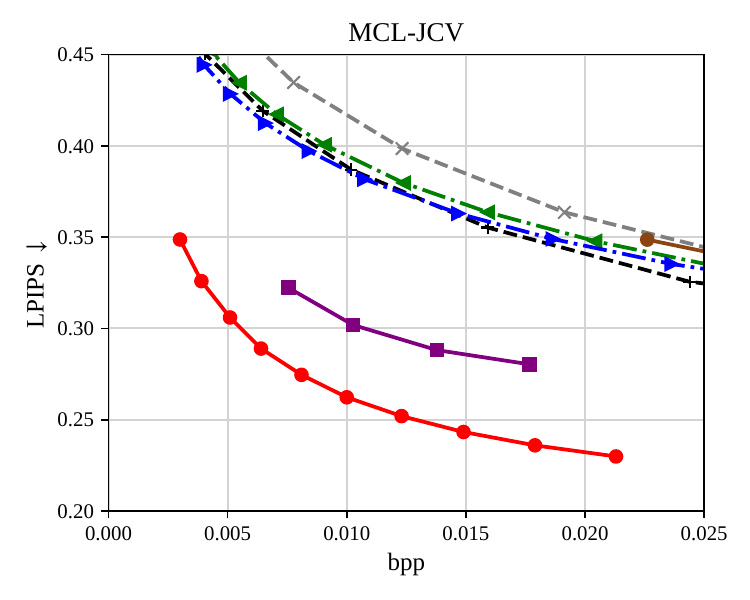}
  \end{subfigure}
  \begin{subfigure}{0.24\textwidth}
    \centering
    \includegraphics[width=\textwidth]{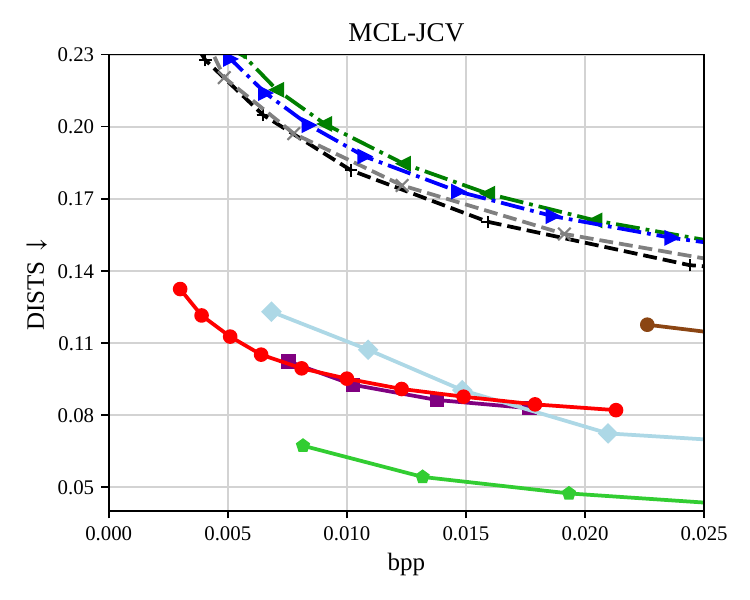}
  \end{subfigure}
  \begin{subfigure}{0.24\textwidth}
    \centering
    \includegraphics[width=\textwidth]{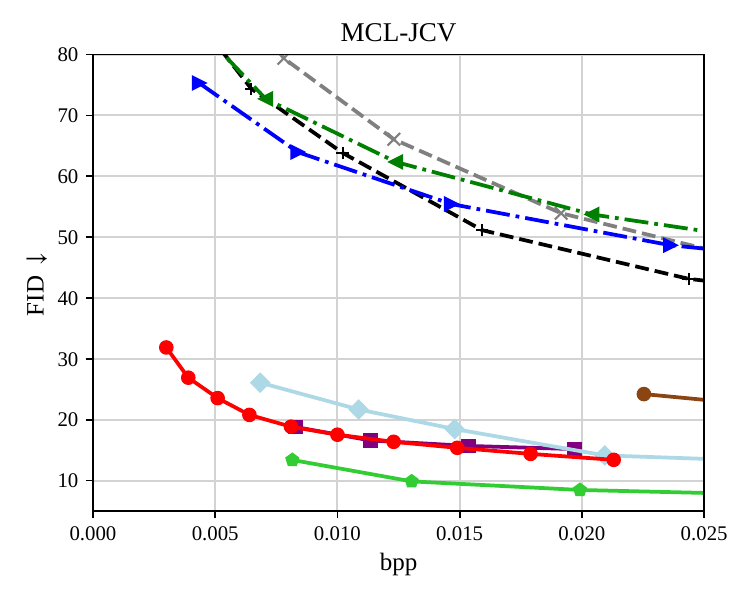}
  \end{subfigure}
  \begin{subfigure}{0.24\textwidth}
    \centering
    \includegraphics[width=\textwidth]{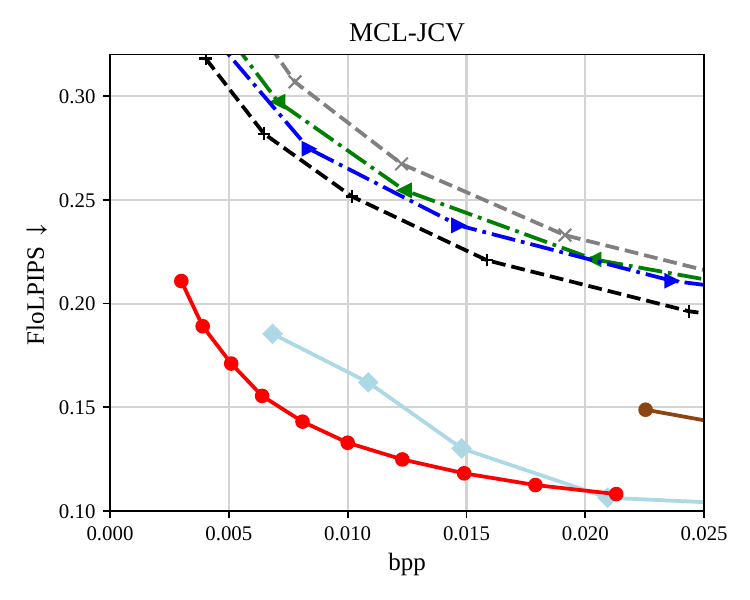}
  \end{subfigure}

  \begin{subfigure}{0.24\textwidth}
    \centering
    \includegraphics[width=\textwidth]{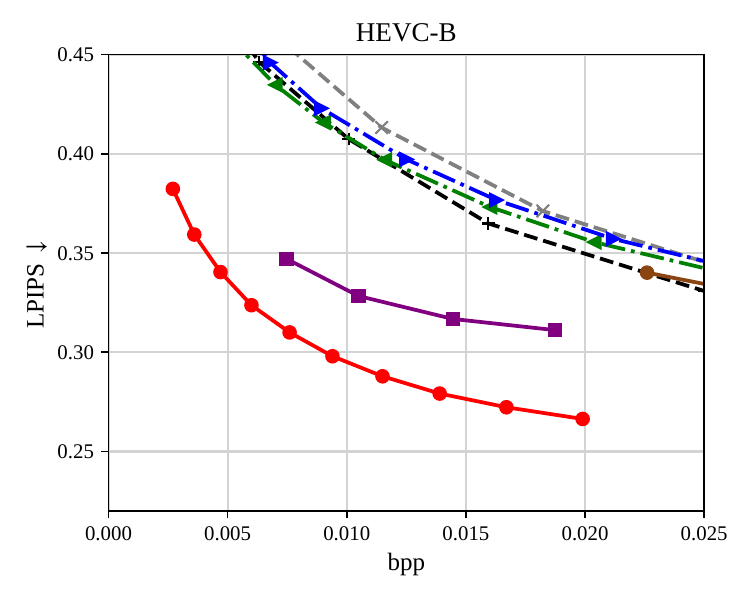}
  \end{subfigure}
  \begin{subfigure}{0.24\textwidth}
    \centering
    \includegraphics[width=\textwidth]{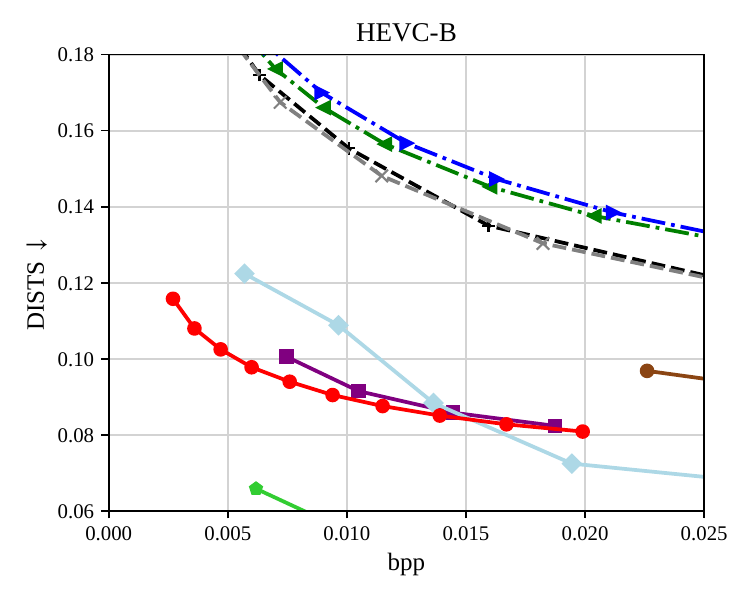}
  \end{subfigure}
  \begin{subfigure}{0.24\textwidth}
    \centering
    \includegraphics[width=\textwidth]{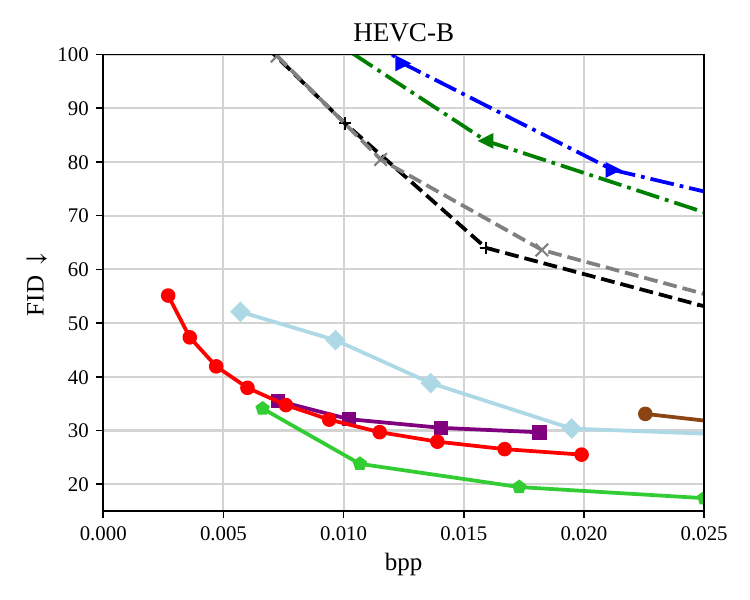}
  \end{subfigure}
  \begin{subfigure}{0.24\textwidth}
    \centering
    \includegraphics[width=\textwidth]{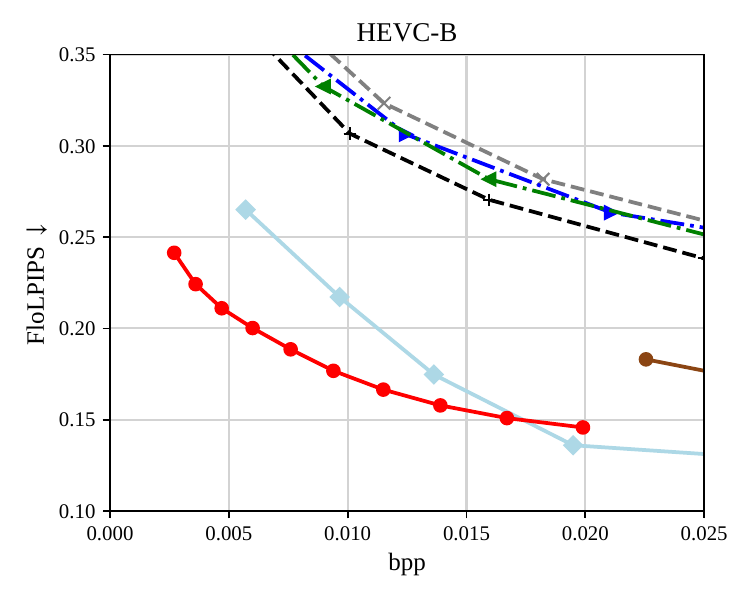}
  \end{subfigure}

  \caption{Rate-perceptual curves comparing GVC-RT with other methods on UVG, MCL-JCV, and HEVC-B using LPIPS, DISTS, FID, and FloLPIPS.}
  \Description{A three-by-four grid of rate-quality plots for UVG, MCL-JCV, and HEVC-B. Columns report LPIPS, DISTS, FID, and FloLPIPS, and each plot compares GVC-RT with traditional, neural, and generative video codecs across bitrates.}
  \label{fig:curves}
\end{figure*}

\section{Experimental Results}

\subsection{Implementation Details}

\noindent\textbf{Datasets.} For training, we use Vimeo-90k~\cite{vimeo90k} to pre-train GVC-RT. To strictly mitigate error propagation during long-sequence inference, we further fine-tune the model using extended 29-frame sequences extracted from the original Vimeo videos~\cite{ori_vimeo}. Comprehensive evaluation is conducted on widely adopted benchmarks, including UVG~\cite{UVG}, HEVC Class B~\cite{flynn16common}, and MCL-JCV~\cite{MCLJCV}.

\noindent\textbf{Training Details.} Our entire model is trained and evaluated directly on raw RGB frames. In Stage I, we pre-train the asymmetric LFQ-AE using individual image frames extracted from the Vimeo-90k dataset~\cite{vimeo90k}, employing randomly cropped $256 \times 256$ spatial patches. To maximize the representation capacity of the generative prior, the LFQ latent dimension $C_l$ is set to 18, corresponding to a codebook size of 262,144. During Stages II and III, the training data transitions to video sequences from the same dataset to effectively model temporal dynamics, while maintaining the consistent $256 \times 256$ spatial cropping. 
To accommodate variable bitrates within a single unified model, we randomly assign a quantization parameter ($qp$) index ranging from 0 to 9 in each training iteration. Following~\cite{DCVCDC}, the corresponding $\lambda$ values are interpolated between 0.08 and 0.9. In a group of 4 pictures, the $qp$ offset is set to [0, 2, 0, 1] for hierarchical quality. We follow~\cite{dcvcfm} to adopt a hierarchical weight setting for the distortion term to support a hierarchical quality structure. Regarding the optimization hyperparameters, the feature alignment weights $w_{cos}$ and $w_{margin}$ in Stage II and III are set to 0.2 and 0.02, respectively. Furthermore, the adversarial loss weight $w_{adv}$ is fixed at 0.1 and remains constant across all three training stages.

\begin{table*}[t] 
\centering
\caption{BD-Rate (\%) comparison of different video compression methods on three datasets. H.266/VVC's reference software VTM-17.0 is used as the anchor. Coding FPS is measured on an NVIDIA RTX 4090 GPU.}
\renewcommand{\arraystretch}{1.0}
\begin{tabular}{lcccccccccc}
\toprule
\multirow{2}{*}{Method} & \multicolumn{2}{c}{HEVC-B} & \multicolumn{2}{c}{MCL-JCV} & \multicolumn{2}{c}{UVG} & \multicolumn{2}{c}{Average} & \multicolumn{2}{c}{Coding FPS $\uparrow$ } \\
\cmidrule(lr){2-3} \cmidrule(lr){4-5} \cmidrule(lr){6-7} \cmidrule(lr){8-9} \cmidrule(lr){10-11}
 & LPIPS $\downarrow$  & DISTS $\downarrow$  & LPIPS $\downarrow$  & DISTS $\downarrow$  & LPIPS $\downarrow$  & DISTS $\downarrow$  & LPIPS $\downarrow$  & DISTS $\downarrow$  & Encoding & Decoding \\
\midrule
HM-16.25 & 31.7 & 5.7 & 47.8 & 10.4 & 27.9 & -0.6 & 35.8 & 5.2 & 0.027 & 12.0 \\
DCVC-FM (half) & 14.0 & 36.2 & 11.6 & 39.0 & 6.7 & 37.3 & 10.8 & 37.5 & 3.7 & 4.4 \\
DCVC-RT (half) & 4.4 & 24.6 & -0.8 & 27.8 & 1.1 & 27.2 & 1.6 & 26.5 & \textbf{123.1} & \textbf{113.0} \\
GLC-Video & -60.8 & -85.5 & -70.9 & \textbf{-93.0} & -40.3 & -91.4 & -57.3 & -90.0 & 6.6 & 3.9 \\
\rowcolor{gray!20}
GVC-RT (half) & \textbf{-78.8} & \textbf{-89.9} & \textbf{-83.3} & -92.1 & \textbf{-80.0} & \textbf{-92.9} & \textbf{-80.7} & \textbf{-91.6} & \textbf{123.1} & 55.1 \\
\bottomrule
\end{tabular}
\label{tab:bd_rate}
\end{table*}

\noindent\textbf{Compared Methods.} We compare our proposed GVC-RT against several SOTA video compression approaches, including traditional codecs (H.265/HEVC HM 16.25~\cite{HM}, H.266/VVC VTM 17.0~\cite{VTM}), MSE-optimized NVCs (DCVC-FM~\cite{dcvcfm}, DCVC-RT~\cite{DCVCRT}), and perceptual-optimized GVCs (PLVC~\cite{PLVC}, GLCVideo~\cite{GLCvideo}). For a fair comparison, we use official implementations or report numerical results directly from the original papers. Since the LFQ prior is pre-trained in RGB space, all baselines are tested under the same RGB mode to ensure a fair comparison protocol. We report all results for the first 96 frames of each video sequence with an intra-period setting of $-1$.

\noindent\textbf{Evaluation Metrics.} We evaluate compression performance using complementary perceptual and distributional metrics. For perceptual quality, we use LPIPS~\cite{lpips} (VGG-based) and DISTS~\cite{dists}, as LPIPS-VGG aligns better with subjective perception in generative compression tasks. Temporal consistency is measured using FloLPIPS~\cite{flolpips}. To quantify distributional differences between reconstructions and ground truth, we compute FID (Fr\'echet Inception Distance)~\cite{fid}. Compression efficiency is quantified by BD-Rate~\cite{bjontegaard2001}.

\begin{figure*}[t]
  \centering
    \includegraphics[width=\linewidth]{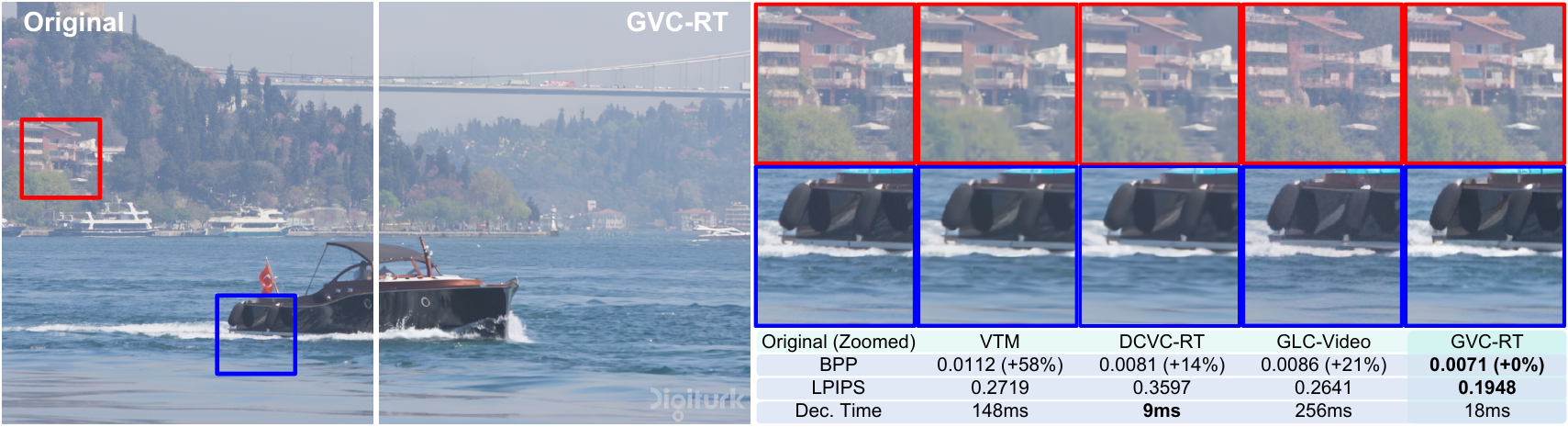}
    \caption{Qualitative comparison across traditional, MSE-optimized, and prior GVC at ultra-low bitrates. GVC-RT preserves finer structures at significantly lower bitrates.} 
  \Description{A grid of video-frame crops compares ground truth, traditional VTM, MSE-optimized DCVC-RT, a prior generative codec, and GVC-RT at ultra-low bitrates. GVC-RT retains sharper window, ship, tree, and texture structures than the baselines.}
  \label{fig:visual}
\end{figure*}

\subsection{Comparison Results}\label{sec:results}

\noindent\textbf{Quantitative Comparisons.} We primarily evaluate perceptual quality using LPIPS, DISTS, FID, and FloLPIPS (Fig.~\ref{fig:curves}), since GVCs are designed for rate-constrained perceptual realism. 

As shown in Fig.~\ref{fig:curves}, our method consistently outperforms existing SOTA GLC-Video across all three datasets, achieving the best scores on all evaluated perceptual metrics. These results demonstrate the superiority of our model in preserving perceptual similarity and texture fidelity (LPIPS/DISTS), while also ensuring distributional consistency and visual realism (FID) with high temporal consistency (FloLPIPS) among these methods.  

In Table~\ref{tab:bd_rate}, we report BD-rate gains for DISTS and LPIPS, together with coding FPS, using VTM-17.0 as the anchor. The tabulated results show that our method substantially outperforms the baselines in perceptual quality, consistently achieving lower BD-Rate values across all these indicators while maintaining the capability of real-time inference.

\noindent\textbf{Qualitative Comparisons.} As shown in Fig.~\ref{fig:visual}, we compare our method with DCVC-RT, a MSE-based NVC, and the traditional VTM-17.0 codec. At low bitrates, both baselines exhibit pronounced blurring artifacts. In contrast, even at lower bitrates, our model better preserves the structure of dynamic content (e.g., the house windows) and maintains textures such as details of the ship and trees, yielding reconstructions that are visually closer to the ground truth.

\definecolor{table}{RGB}{127,127,127}

\begin{table}[t]
\centering
\caption{Ablation study on HEVC-B, with full GVC-RT as the BD-rate anchor. Lower BD-rate is better; decoding FPS is measured on an NVIDIA RTX 4090 GPU.}\label{tab:ablation}
\small
\setlength{\tabcolsep}{3.pt}
\renewcommand{\arraystretch}{0.91}
\begin{tabular}{@{}lcccc@{}}
\toprule
\multirow{2}{*}{Variant} &
\multicolumn{3}{c}{BD-Rate (\%) $\downarrow$} &
\multirow{2}{*}{Dec. FPS $\uparrow$} \\

        \cmidrule(lr){2-4}
        & LPIPS & DISTS & PSNR &\\
\midrule
\multicolumn{5}{l}{\textcolor{table}{\textit{Ablations on Training Stages and Modules}}}  \\
w/o Frozen De-Tokenizer (Stage II)                             & 19.0\% & -3.8\% & 34.4\% &  55.0  \\ 
w/o Joint Fine-tuning (Stage III)                    & 21.7\% & 5.1\% & 62.8\% &  55.0  \\ 
Deeper De-Tokenizer                                 & -13.4\% & -1.9\% & -18.0\% & 44.1  \\ 
\midrule
\multicolumn{5}{l}{\textcolor{table}{\textit{Ablations on Loss Functions and Hyperparameters}}}  \\
BCE + MSE + $\mathcal{L}_{rec}$ (Stage II \& III)   & 225.3\% & N/A & 17.1\% &  55.0  \\ 
w/o $\mathcal{L}_{margin}$ (Stage II \& III)                 & 143.1\% & 250.9\% & -5.9\% & 55.0  \\ 
$w_{adv}=0.5$ (Loss weight)                                                & 11.1\% & 2.2\% & 55.6\% &  55.0  \\ 
$w_{LPIPS}=2.0$ (Loss weight)                                              & 13.3\% & 6.9\% & 79.4\% &  55.0  \\ 
\bottomrule
\end{tabular}
\end{table}

\noindent\textbf{Complexity Analysis.} Table~\ref{tab:bd_rate} reports all coding speeds on an NVIDIA RTX 4090 GPU. GVC-RT matches the 123.1 FPS encoding speed of DCVC-RT~\cite{DCVCRT} and reaches 55.1 FPS decoding, compared with 3.9 FPS for GLC-Video~\cite{GLCvideo}. This is a remarkable decoding speedup of approximately 14× and supports real-time 1080p inference on a consumer-grade GPU.

\noindent\textbf{Ablation Study.} Table~\ref{tab:ablation} evaluates the contributions of the training strategy, de-tokenizer architecture, and loss design. Disabling joint fine-tuning consistently worsens BD-rate performance across all three metrics, demonstrating the importance of end-to-end adaptation. Freezing Stage II yields a marginal DISTS improvement but degrades LPIPS and PSNR. Increasing the de-tokenizer depth improves all three BD-rate results at the cost of reducing the decoding speed from 55.0 to 44.1 FPS, revealing a clear quality--efficiency trade-off. Among the loss variants, replacing the objectives used in Stages II and III causes the largest LPIPS BD-rate degradation, while removing $\mathcal{L}_{margin}$ also impairs perceptual quality. Moreover, increasing either reported loss weight worsens all three BD-rates, supporting the adopted loss configuration.

\section{Conclusion}

In this paper, we have proposed GVC-RT, a novel GVC framework that successfully bridges the critical gap between high-realism ultra-low bitrate compression and real-time processing capability. By systematically analyzing the latency bottlenecks of existing tokenizer-based generative codecs, we have designed a highly efficient asymmetric architecture. GVC-RT offloads the complex generative space alignment to the training phase and directly learns a soft-LFQ latent distribution, completely bypassing computationally expensive tokenization and diffusion during inference. Extensive evaluations demonstrate that our method achieves competitive perceptual quality at extreme compression ratios, while delivering 1080p real-time processing speeds on an NVIDIA RTX 3090 GPU, paving the way for the practical deployment of GVCs.

\begin{acks}
This work was supported in part by the National Key Research and Development Program of China under Grant 2024YFF0509700, in part by the National Natural Science Foundation of China under Grant 62371063, Grant 62471290, Grant 62321001, in part by the Beijing Municipal Natural Science Foundation under Grant L232047, in part by Postdoctoral Fellowship Program of CPSF under Grant Number GZB20250810, in part by the China Postdoctoral Science Foundation under Grant Number 2025M783515, in part by the Beijing Nova Program under Grant Number 20250484834, and in part by the Fundamental Research Funds for the Central Universities under Grant 2025TSQY11.
\end{acks}

\bibliographystyle{ACM-Reference-Format}
\bibliography{reference}

@String{Computing = "Computing" }

@String{Computer = "{IEEE} Computer" }

@String{Springer = "Springer-Verlag" }

@inproceedings{zhang2025ultra,
  title={Ultra-Low Bitrate Perceptual Image Compression with Shallow Encoder},
  author={Zhang, Tianyu and Liu, Dong and Chen, Chang Wen},
  booktitle={Proceedings of the IEEE/CVF Conference on Computer Vision and Pattern Recognition (CVPR)},
  pages={12118--12128},
  month={June},
  year={2026},
  location={Denver, CO, USA},
  publisher={IEEE},
  address={Piscataway, NJ, USA}
}

@inproceedings{LMBeatsDiff,
  title={Language Model Beats Diffusion - Tokenizer is key to visual generation},
  author={Yu, Lijun and Lezama, Jose and Gundavarapu, Nitesh Bharadwaj and Versari, Luca and Sohn, Kihyuk and Minnen, David and Cheng, Yong and Gupta, Agrim and Gu, Xiuye and Hauptmann, Alexander G and Gong, Boqing and Yang, Ming-Hsuan and Essa, Irfan and Ross, David A and Jiang, Lu},
  booktitle={The Twelfth International Conference on Learning Representations},
  year={2024},
  numpages={19},
  location={Vienna, Austria},
  publisher={OpenReview.net},
  address={Amherst, MA, USA},
  url={https://openreview.net/forum?id=gzqrANCF4g},
  lastaccessed={August 3, 2026}
}

@misc{luo2024open,
  title={{Open-MAGVIT2}: An Open-Source Project Toward Democratizing Auto-Regressive Visual Generation},
  author={Luo, Zhuoyan and Shi, Fengyuan and Ge, Yixiao and Yang, Yujiu and Wang, Limin and Shan, Ying},
  year={2024},
  eprinttype={arXiv},
  eprint={2409.04410},
  eprintclass={cs.CV}
}

@inproceedings{FSQ,
  title={Finite Scalar Quantization: {VQ}-{VAE} Made Simple},
  author={Mentzer, Fabian and Minnen, David and Agustsson, Eirikur and Tschannen, Michael},
  booktitle={The Twelfth International Conference on Learning Representations},
  year={2024},
  numpages={12},
  location={Vienna, Austria},
  publisher={OpenReview.net},
  address={Amherst, MA, USA},
  url={https://openreview.net/forum?id=8ishA3LxN8},
  lastaccessed={August 3, 2026}
}

@inproceedings{DCVCRT,
  title={Towards practical real-time neural video compression},
  author={Jia, Zhaoyang and Li, Bin and Li, Jiahao and Xie, Wenxuan and Qi, Linfeng and Li, Houqiang and Lu, Yan},
  booktitle={Proceedings of the IEEE/CVF Conference on Computer Vision and Pattern Recognition (CVPR)},
  pages={12543--12552},
  month={June},
  year={2025},
  location={Nashville, TN, USA},
  publisher={IEEE},
  address={Piscataway, NJ, USA},
  doi={10.1109/CVPR52734.2025.01170}
}

@inproceedings{patchgan,
  author={Isola, Phillip and Zhu, Jun-Yan and Zhou, Tinghui and Efros, Alexei A.},
  title={Image-to-Image Translation with Conditional Adversarial Networks},
  booktitle={Proceedings of the IEEE Conference on Computer Vision and Pattern Recognition (CVPR)},
  pages={1125--1134},
  month={July},
  year={2017},
  location={Honolulu, HI, USA},
  publisher={IEEE},
  address={Piscataway, NJ, USA},
  doi={10.1109/CVPR.2017.632}
}

@misc{diffusers,
  author = {von Platen, Patrick and Patil, Suraj and Lozhkov, Anton and Cuenca, Pedro and Lambert, Nathan and Rasul, Kashif and Davaadorj, Mishig and Nair, Dhruv and Paul, Sayak and Berman, William and Xu, Yiyi and Liu, Steven and Wolf, Thomas},
  title = {Diffusers: State-of-the-art diffusion models},
  year = {2022},
  howpublished = {GitHub repository},
  url = {https://github.com/huggingface/diffusers},
  lastaccessed = {August 3, 2026}
}

@InProceedings{DCB,
  author = {Chollet, Francois},
  title = {Xception: Deep Learning With Depthwise Separable Convolutions},
  booktitle = {Proceedings of the IEEE Conference on Computer Vision and Pattern Recognition (CVPR)},
  month = {July},
  year = {2017},
  pages = {1251--1258},
  location = {Honolulu, HI, USA},
  publisher = {IEEE},
  address = {Piscataway, NJ, USA},
  doi = {10.1109/CVPR.2017.195}
}

@inproceedings{PLVC,
  title={Perceptual Learned Video Compression with Recurrent Conditional {GAN}},
  author={Yang, Ren and Timofte, Radu and Van Gool, Luc},
  booktitle={Proceedings of the International Joint Conference on Artificial Intelligence},
  pages={1537--1544},
  year={2022},
  location={Vienna, Austria},
  publisher={International Joint Conferences on Artificial Intelligence Organization},
  address={Menlo Park, CA, USA},
  doi={10.24963/ijcai.2022/214}
}

@article{GLCvideo,
  title={Generative latent coding for ultra-low bitrate image and video compression},
  author={Qi, Linfeng and Jia, Zhaoyang and Li, Jiahao and Li, Bin and Li, Houqiang and Lu, Yan},
  journal={IEEE Transactions on Circuits and Systems for Video Technology},
  volume={35},
  number={10},
  pages={10500--10515},
  year={2025},
  doi={10.1109/TCSVT.2025.3571944}
}

@article{diffusion,
  title={Denoising diffusion probabilistic models},
  author={Ho, Jonathan and Jain, Ajay and Abbeel, Pieter},
  journal={Advances in Neural Information Processing Systems},
  volume={33},
  pages={6840--6851},
  year={2020}
}

@article{DiffVC,
  title={Diffusion-based perceptual neural video compression with temporal diffusion information reuse},
  author={Ma, Wenzhuo and Chen, Zhenzhong},
  journal={ACM Transactions on Multimedia Computing, Communications, and Applications},
  volume={21},
  number={12},
  articleno={345},
  numpages={22},
  year={2025},
  doi={10.1145/3761815}
}

@inproceedings{DiffVC_OSD,
  title={DiffVC-OSD: One-Step Diffusion-based Perceptual Neural Video Compression Framework},
  author={Ma, Wenzhuo and Chen, Zhenzhong},
  booktitle={2025 International Conference on Visual Communications and Image Processing (VCIP)},
  pages={1--5},
  year={2025},
  location={Klagenfurt, Austria},
  publisher={IEEE},
  address={Piscataway, NJ, USA},
  doi={10.1109/VCIP67698.2025.11396819}
}

@inproceedings{fid,
  title={GANs trained by a two time-scale update rule converge to a local nash equilibrium},
  author={Heusel, Martin and Ramsauer, Hubert and Unterthiner, Thomas and Nessler, Bernhard and Hochreiter, Sepp},
  booktitle={Proceedings of the 31st International Conference on Neural Information Processing Systems},
  pages={6629--6640},
  year={2017},
  location={Long Beach, CA, USA},
  publisher={Curran Associates, Inc.},
  address={Red Hook, NY, USA}
}

@misc{DiffVC_RT,
  title={DiffVC-RT: Towards Practical Real-Time Diffusion-based Perceptual Neural Video Compression},
  author={Ma, Wenzhuo and Chen, Zhenzhong},
  year={2026},
  eprinttype={arXiv},
  eprint={2601.20564},
  eprintclass={cs.CV}
}

@misc{yoda,
  title={{YODA}: Yet Another One-step Diffusion-based Video Compressor},
  author={Li, Xingchen and Zhang, Junzhe and Shi, Junqi and Lu, Ming and Ma, Zhan},
  year={2026},
  howpublished={IEEE Transactions on Circuits and Systems for Video Technology, Early Access},
  doi={10.1109/TCSVT.2026.3714453}
}

@inproceedings{gnvc,
  title={Generative Neural Video Compression via Video Diffusion Prior},
  author={Mao, Qi and Cheng, Hao and Yang, Tinghan and Jin, Libiao and Ma, Siwei},
  booktitle={Proceedings of the IEEE/CVF Conference on Computer Vision and Pattern Recognition (CVPR)},
  pages={43239--43248},
  month={June},
  year={2026},
  location={Denver, CO, USA},
  publisher={IEEE},
  address={Piscataway, NJ, USA}
}

@inproceedings{s2vc,
  title={Single-step Diffusion-based Video Coding with Semantic-Temporal Guidance},
  author={Xue, Naifu and Jia, Zhaoyang and Li, Jiahao and Li, Bin and Zheng, Zihan and Zhang, Yuan and Lu, Yan},
  booktitle={Proceedings of the IEEE/CVF Conference on Computer Vision and Pattern Recognition (CVPR)},
  pages={9752--9761},
  month={June},
  year={2026},
  location={Denver, CO, USA},
  publisher={IEEE},
  address={Piscataway, NJ, USA}
}

@misc{ori_vimeo,
  author={Xue, Tianfan and Chen, Baian and Wu, Jiajun and Wei, Donglai and Freeman, William T.},
  title={Original Vimeo Links for the {Vimeo-90K} Dataset},
  year={2019},
  url={https://github.com/anchen1011/toflow/blob/master/data/original_vimeo_links.txt},
  lastaccessed={April 2, 2026}
}

@inproceedings{lpips,
  title={The unreasonable effectiveness of deep features as a perceptual metric},
  author={Zhang, Richard and Isola, Phillip and Efros, Alexei A and Shechtman, Eli and Wang, Oliver},
  booktitle={Proceedings of the IEEE/CVF Conference on Computer Vision and Pattern Recognition},
  pages={586--595},
  year={2018},
  location={Salt Lake City, UT, USA},
  publisher={IEEE},
  address={Piscataway, NJ, USA},
  doi={10.1109/CVPR.2018.00068}
}

@inproceedings{vgg,
  title={Very deep convolutional networks for large-scale image recognition},
  author={Simonyan, Karen and Zisserman, Andrew},
  booktitle={International Conference on Learning Representations (ICLR)},
  year={2015},
  numpages={14},
  location={San Diego, CA, USA},
  publisher={International Conference on Learning Representations},
  address={San Diego, CA, USA}
}

@article{dists,
  title={Image quality assessment: Unifying structure and texture similarity},
  author={Ding, Keyan and Ma, Kede and Wang, Shiqi and Simoncelli, Eero P},
  journal={IEEE Transactions on Pattern Analysis and Machine Intelligence},
  volume={44},
  number={5},
  pages={2567--2581},
  year={2022},
  doi={10.1109/TPAMI.2020.3045810}
}

@inproceedings{dcvcfm,
  title={Neural video compression with feature modulation},
  author={Li, Jiahao and Li, Bin and Lu, Yan},
  booktitle={Proceedings of the IEEE/CVF Conference on Computer Vision and Pattern Recognition},
  pages={26099--26108},
  year={2024},
  location={Seattle, WA, USA},
  publisher={IEEE},
  address={Piscataway, NJ, USA},
  doi={10.1109/CVPR52733.2024.02466}
}

@article{vvc,
  title={Overview of the versatile video coding ({VVC}) standard and its applications},
  author={Bross, Benjamin and Wang, Ye-Kui and Ye, Yan and Liu, Shan and Chen, Jianle and Sullivan, Gary J and Ohm, Jens-Rainer},
  journal={IEEE Transactions on Circuits and Systems for Video Technology},
  volume={31},
  number={10},
  pages={3736--3764},
  year={2021},
  publisher={IEEE}
}

@inproceedings{fvc,
  title={{FVC}: A new framework towards deep video compression in feature space},
  author={Hu, Zhihao and Lu, Guo and Xu, Dong},
  booktitle={Proceedings of the IEEE/CVF Conference on Computer Vision and Pattern Recognition},
  pages={1502--1511},
  year={2021},
  location={Nashville, TN, USA},
  publisher={IEEE},
  address={Piscataway, NJ, USA},
  doi={10.1109/CVPR46437.2021.00155}
}

@article{li2021dcvc,
  title={Deep contextual video compression},
  author={Li, Jiahao and Li, Bin and Lu, Yan},
  journal={Advances in Neural Information Processing Systems},
  volume={34},
  pages={18114--18125},
  year={2021}
}

@inproceedings{dhvc,
  title={Deep hierarchical video compression},
  author={Lu, Ming and Duan, Zhihao and Zhu, Fengqing and Ma, Zhan},
  booktitle={Proceedings of the AAAI Conference on Artificial Intelligence},
  volume={38},
  pages={8859--8867},
  year={2024},
  location={Vancouver, BC, Canada},
  publisher={Association for the Advancement of Artificial Intelligence},
  address={Washington, DC, USA},
  doi={10.1609/aaai.v38i8.28733}
}

@inproceedings{UVG,
  title={{UVG} dataset: 50/120fps 4{K} sequences for video codec analysis and development},
  author={Mercat, Alexandre and Viitanen, Marko and Vanne, Jarno},
  booktitle={Proceedings of the 11th ACM Multimedia Systems Conference},
  pages={297--302},
  year={2020},
  location={Istanbul, Turkey},
  publisher={Association for Computing Machinery},
  address={New York, NY, USA},
  doi={10.1145/3339825.3394937}
}

@article{vimeo90k,
  title={Video enhancement with task-oriented flow},
  author={Xue, Tianfan and Chen, Baian and Wu, Jiajun and Wei, Donglai and Freeman, William T},
  journal={International Journal of Computer Vision},
  volume={127},
  number={8},
  pages={1106--1125},
  year={2019},
  publisher={Springer}
}

@inproceedings{MCLJCV,
  title={{MCL-JCV}: A {JND}-based {H.264/AVC} Video Quality Assessment Dataset},
  author={Wang, Haiqiang and Gan, Weihao and Hu, Sudeng and Lin, Joe Yuchieh and Jin, Lina and Song, Longguang and Wang, Ping and Katsavounidis, Ioannis and Aaron, Anne and Kuo, C-C Jay},
  booktitle={2016 IEEE International Conference on Image Processing (ICIP)},
  pages={1509--1513},
  year={2016},
  location={Phoenix, AZ, USA},
  publisher={IEEE},
  address={Piscataway, NJ, USA},
  doi={10.1109/ICIP.2016.7532610}
}

@inproceedings{flolpips,
  author    = {Danier, Duolikun and Zhang, Fan and Bull, David},
  title     = {FloLPIPS: A Bespoke Video Quality Metric for Frame Interpolation},
  booktitle = {2022 Picture Coding Symposium (PCS)},
  year      = {2022},
  pages     = {283--287},
  location  = {San Jose, CA, USA},
  publisher = {IEEE},
  address   = {Piscataway, NJ, USA},
  doi       = {10.1109/PCS56426.2022.10018062}
}

@inproceedings{DCVCDC,
  title={Neural video compression with diverse contexts},
  author={Li, Jiahao and Li, Bin and Lu, Yan},
  booktitle={Proceedings of the IEEE/CVF Conference on Computer Vision and Pattern Recognition},
  pages={22616--22626},
  year={2023},
  location={Vancouver, BC, Canada},
  publisher={IEEE},
  address={Piscataway, NJ, USA},
  doi={10.1109/CVPR52729.2023.02166}
}

@misc{HM,
  title = {{HEVC} Reference Software {HM-16.25}},
  author = {{Joint Video Experts Team (JVET)}},
  year = {2022},
  url = {https://vcgit.hhi.fraunhofer.de/jvet/HM/-/tags/HM-16.25},
  lastaccessed = {August 1, 2026}
}

@techreport{flynn16common,
  author={Bossen, Frank},
  title={Common Test Conditions and Software Reference Configurations},
  institution={Joint Collaborative Team on Video Coding (JCT-VC)},
  number={JCTVC-L1100},
  address={Geneva, Switzerland},
  year={2013}
}

@misc{VTM,
  title = {{VVC} Reference Software {VTM-17.0}},
  author = {{Joint Video Experts Team (JVET), Fraunhofer HHI}},
  year = {2022},
  url = {https://vcgit.hhi.fraunhofer.de/jvet/VVCSoftware_VTM/-/tags/VTM-17.0},
  lastaccessed = {August 1, 2026}
}

@InProceedings{Lu_2019_CVPR,
  author = {Lu, Guo and Ouyang, Wanli and Xu, Dong and Zhang, Xiaoyun and Cai, Chunlei and Gao, Zhiyong},
  title = {DVC: An End-To-End Deep Video Compression Framework},
  booktitle = {Proceedings of the IEEE/CVF Conference on Computer Vision and Pattern Recognition (CVPR)},
  month = {June},
  year = {2019},
  pages = {11006--11015},
  location = {Long Beach, CA, USA},
  publisher = {IEEE},
  address = {Piscataway, NJ, USA},
  doi = {10.1109/CVPR.2019.01126}
}

@inproceedings{MS-SSIM,
  title={Multiscale structural similarity for image quality assessment},
  author={Wang, Zhou and Simoncelli, Eero P and Bovik, Alan C},
  booktitle={Conference Record of the Thirty-Seventh Asilomar Conference on Signals, Systems and Computers},
  volume={2},
  pages={1398--1402},
  year={2003},
  location={Pacific Grove, CA, USA},
  publisher={IEEE},
  address={Piscataway, NJ, USA},
  doi={10.1109/ACSSC.2003.1292216}
}

@techreport{bjontegaard2001,
  author={Bj{\o}ntegaard, Gisle},
  title={Calculation of Average {PSNR} Differences between {RD}-Curves},
  institution={ITU-T Video Coding Experts Group (VCEG)},
  number={VCEG-M33},
  address={Austin, Texas, USA},
  month={April},
  year={2001}
}

\end{document}